\documentclass[aps,twocolumn,nofootinbib]{revtex4}

\usepackage{amsmath,amssymb}
\usepackage{epsfig}
\usepackage{color}
\newcommand{\n}{\nonumber}
\newcommand{\be}{\begin{equation}}
\newcommand{\ee}{\end{equation}}
\newcommand{\bea}{\begin{eqnarray}}
\newcommand{\eea}{\end{eqnarray}}

\usepackage{color}

\begin{document}

\title{Dirac's variational approach to semiclassical Kramers problem in Smoluchowski limit}
%: Dirac's variational principle}

\author{Choon-Lin Ho
%\footnote{email: hcl@mail.tku.edu.tw
%Tel: +886-2-26215656 Ext 2521\\
%Fax: +886-2-26209917}
}
\affiliation{Department of Physics, Tamkang University, Tamsui
25137, Taiwan}

%\date{28 Jun 2017} 14 Apr 2017}, % 5 Apr 2017}    %4 Jan 2017}
\begin{abstract}

Kramers escape from a metastable state in the presence of both thermal and quantum fluctuations under strong damping is treated as a  thermally activated process in a quantum modified semiclassical potential.    Dirac's time-dependent variational method together with the Jackiw-Kerman function is employed to derive the semiclassical potential.  Quantum correction is incorporated in the drift potential, and is determined by quasi-stationary conditions and minimal uncertainty relation.  The semiclassical rate obtained here is consistent in form with those from the quantum Smoluchowski equations  deduced  heuristically  by modifying the diffusion coefficient  using the path-integral method. Unlike approaches using the path-integral, which involves continuation into imaginary time, the approach here is simpler and more easily understood in terms of classical picture. 

\end{abstract}

%\pacs{03.65.Sq,05.10.Gg,72.70.+m}
% 03.65.Sq  Semiclassical theories and applications
%  05.10.Gg Stochastic analysis methods (Fokker-Planck, Langevin, etc.)
%  72.70.+m  Noise processes and phenomena

%\keywords{Kramers problem, Smoluchowski equation, semiclassical approach, variational principle}

\maketitle

%%%%%%%%%%%%%%%%%%%%%%%%%%
\section{Introduction}

The problem of escape from metastable states is ubiquitous in physics, chemistry, and biology. 
A metastable state is a temporary stable non-equilibrium state of a system with a considerable long lifetime before the particle transits to a lower energy state.  Common examples are the alpha-decay of radioactive atoms, supersaturated vapor,  transition states in chemical reactions, etc. System in such a state can be considered as being trapped in a local minimum of the energy potential of the system. Escape from a metastable state can be triggered by thermal fluctuations, or quantum tunneling, or  both effects.  

The work of Kramers \cite{K} represents a classic analysis of such problems as thermally activated processes.  He treated the escape process as a Brownian motion of the particle in the metastable state driven by thermal noise. The dynamics was given in terms of some probability density function that satisfies an equation of the Fokker-Planck type, called the Kramers equation \cite{G,R}.

In recent years there has been much interest in the study of quantum Kramers problem, which considers both the thermal and quantum fluctuations.  This is mainly motivated by the work of Caldeira and Leggett \cite{M1}.  They studied quantum tunneling in the presence of dissipation  at zero temperature. 
 Their motivation was to understand quantum mechanics at the macroscopic level as stimulated by the problem of Schr\"odinger cat.  The method of path-integral was found useful in their work, and has since become the main tool in subsequent works by others.   Extension to finite temperatures were given in \cite{M2,M3,M4}.  Candidates for such macroscopic quantum phenomena include, for example, Josephson junction and SQUID circuit, single-domain magnet, liquid Helium-3 and -4 mixture \cite{M5},  and certain phenomena in relativistic physics \cite{M6}.  For reviews of the earlier works, see e.g., \cite{QK1,QK2}. Soon after, different approaches have been proposed to study the problem, e.g., by Wigner phase space \cite{W1,W2}, and by quantum Smoluchowski equations \cite{A1,A2,A3,S1,S2}.  
  
 For quantum Kramers problem, one is basically interested in the quantum correction to the thermally activated escape rate over a potential barrier.  In this case, a semiclassical approach is usually adopted. 
Most of the works appear so far make use of the complicated machinery of path integral. This method is elegant and powerful in dealing with open systems with noise and dissipation.   However, it involves continuation into imaginary time, which in our opinion does not provide a direct and  intuitive understanding of the physical process involved.  Furthermore, unlike the classical theory of Kramers, which considers the evolution of probability distribution function of the Brownian motion of a particle under a force field, the path integral method deals mainly with the evaluation of quantum partition function of the system interacting with a bath of harmonic oscillators having a certain distribution of frequencies. Different distributions represent different damping characteristics. Thus the result depends very much on the frequency spectrum  of the  oscillators assumed. 

In our view, a simple and more intuitive semiclassical approach has been the time-dependent variational method initiated by Dirac \cite{D1,D2}. This method gives the effective semiclassical Hamiltonian very directly.
 Previously we have employed this method to study the phases of the ground states of the quantum Frankel-Kontorova model \cite{H1}, and the quantum chaotic behavior of the H\'enon-Heiles model \cite{H2} (a different chaotic system has been considered by this method in \cite{E}).  In these work we found it very useful to  take the Jackiw-Kerman function \cite{JK,C} as the trial wavefuncrtion for variational method.

Here  we would like to apply this approach  to study the semiclassical Kramers escape problem.  The aim is to incorporate quantum effect in the semiclassical way, i.e., to see how quantum effect modifies the potential that trap the metastable state.  This amounts to obtaining an effective potential with quantum correction. 
The escape rate from this new potential is then calculated following Kramers' original method.  

%%%%%%%%%%%%%%
\section{Classical Kramers escape rate}

A particle of mass $M$ moving in a potential $V(x)$ together with a damping force and a stochastic force is governed by the stochastic equation
\be
M \ddot{x}(t) + M\gamma \dot{x}(t) + V^\prime (x) =M\gamma\,\xi(t).
\label{GLE}
\ee
Here the ``dot" and the ``prime" represent derivatives with respect to time and space, respectively.
$-M\gamma \dot{x}$ is the damping force with damping rate per unit mass $\gamma$. The stochastic force $\xi(t)$ depends only on the time $t$ and obeys the stationary Gaussian statistics

\be 
\langle \xi(t)\rangle =0,~~ \langle \xi(t)\xi(0)\rangle= 2 D \delta(t),~~~D\equiv \frac{k_B T}{M\gamma},
\ee
where $k_B$ is the Boltzmann constant and $T$ the temperature.

%-- strong damping
For strong damping, one can ignore the first term in (\ref{GLE}) and obtain the Langevin equation
\be
\dot{x}(t) + \frac{V^\prime (x)}{M\gamma} =\xi(t).
\label{LE}
\ee
By It\^o's  prescription, the Fokker-Planck equation corresponding to (\ref{LE}) is the Smoluchowski equation 
\be
\frac{\partial }{\partial t}P(x,t)= \frac{\partial}{\partial x}\left(\frac{V^\prime (x)}{M\gamma}\,P(x,t)\right)
+ D  \frac{\partial^2}{\partial x^2}\,P(x,t).
\label{S}
\ee
Here $P(x,t)$ is the probability density function describing the motion of the particle, and 
$V^\prime (x)/M\gamma$  and $D$ are the drift and the diffusion coefficient, respectively.

For Kramers escape problem, it is typical to consider  a potential with two minima separated by a barrier, such as the asymmetric/symmetric  double-well potential, or a completely asymmetric potential with a minimum and a maximum beyond which the potential is non-confining. 
Here, for definiteness, we shall consider  a potential $V(x)$ having a minimum $V(x_0)=0$ at the origin $x_0=0$, a maximum with barrier height $V(x_b)>0$ at $x_b>0$, and a local minimum at $x_m>x_b$ (such as  the double-well),   or $V(x)\to V_\infty<0$ as $x\to \infty$.  Without loss of generality, we assume the boundary condition $V(x)\to \infty$ as $x\to -\infty$, i.e., a reflecting boundary at $x\to -\infty$.  The particle, once escape from the minimum at $x=0$ , may stay in the minimum at $x_m$ (in this case the particle may escape back to $x_0$ again), or will move towards $x\to +\infty$ without returning.  The double-well is described by a quartic function,  such as  $V(x)=A x^2 (x-B)^2 (A, B>0$).   For the non-confining potential a commonly employed one is the cubic potential, $V(x)=Ax^2-Bx^3 (A, B>0)$.  We consider the situation where $V(x_b)\gg k_B T \gg \hbar \omega_0$, here 
$\omega_0\equiv \sqrt{V^{\prime\prime}(x_0)/M}$ is the frequency of small oscillation around the minimum $V(x_0)$. 

By taking quadratic approximation at the minimal and maximal points, Kramers obtains the classical escape rate \cite{K,G,R}
\be
r_{c} =\frac{\omega_0\omega_b}{2\pi\gamma}\,e^{-\beta [V(x_b)-V(x_0)]}
\label{r_c}
\ee
for the over-damped case $\gamma \gg \omega_b$,  where $\omega_b\equiv \sqrt{|V^{\prime\prime}(x_b)|/M}$ is the frequency around $x_b$ of the inverted potential. 

We would like to see how quantum fluctuations would modify the above escape rate. 
To account for quantum fluctuations, we consider a semiclassical model of the escape problem. 
We modify the potential $V(x)$ to an effective semiclassical potential ${\hat V}(x)$ that incorporates quantum effects. The motion of the escaping particle is then considered to be caused by the extra stochastic noise in this new potential.  To derive the effective semiclassical potential, we adopt Dirac's time-dependent variational
principle \cite{D1} together with the Jackiw-Kerman (JK) function \cite{JK}.  The essence of this approach we shall briefly review in the next section. 

%%%%%%%%%%

\section{Dirac's time-dependent variational approach}  

Quantum Hamiltonian $H$ of a particle of mass $M$ in the potential $V(Q)$ is 
$H=\frac{P^2}{2M} + V(Q)$, where $P, Q$ are the momentum and position operators.

To derive the semiclassical Hamiltonian, 
we
adopt the time-dependent variational principle pioneered by Dirac
\cite{D1,D2}. In this approach, one first constructs the effective action
$\Gamma=\int dt
~\langle \Psi,t|i\hbar\partial_t -{\cal H}|\Psi,t\rangle$ for the quantum system 
described
by $H$ and the time-dependent states $|\Psi,t\rangle$.  Variation of $\Gamma$ is then the  quantum
analogue of the Hamilton's principle.  Of various possible choices of the trial wavefunction of the state of
the quantum system, we find it  simple and elegant to adopt the JK wavefunction \cite{JK}:
\begin{eqnarray}
&&\langle Q|\Psi,t\rangle=\frac{1}{(2\pi\hbar
G)^{1/4}}  \label{JK} 
 \\  && \times % arXiv, 
  \exp\Biggl\{
-\frac{1}{2\hbar}\left(Q-x\right)^2\Bigl[\frac{1}{2}G^{-1}
%\nonumber\\
-2i \Pi\Bigr]+\frac{i}{\hbar}p\left(Q-x\right)
\Biggr\}.\n % for arXiv
% \Biggr\}.
\end{eqnarray}
The real quantities $x(t)$, $p(t)$, $G(t)$ and $\Pi (t)$ are
variational parameters the variations of which at $t=\pm\infty$ are assumed to
vanish. We prefer to use the JK form since the physical meanings of the variational
parameters contained in the JK wavefunction are most transparent, as will be evident
from the discussion below.  

The effective action $\Gamma$ for
the Hamiltonian $H$ can be worked
out to be \cite{C,H1,H2}
\be
\Gamma (x,p,G,\Pi)=
\int dt ~\left[
(p\dot{x}+\hbar\Pi
\dot{G})-H_{sc}\right],  
\label{action}
\ee
where $H_{sc}=\langle\Psi|H|\Psi\rangle$ is the semiclassical
Hamiltonian evaluated to be, up to $\hbar$-terms, 
\bea
&& ~~H_{sc}=
\frac{1}{2M}p^2+ V_{sc}(x),\label{Vsc} \\
&&V_{sc}(x)=V(x)+\hbar\left[\frac{1}{8MG} 
+ \frac{2}{M} G\Pi^2 + \frac12 G V^{\prime\prime}\right].
\n
\eea
 One sees from the form of the effective action $\Gamma$
that $\Pi$ is the canonical conjugate of $\hbar G$.
 The second line of (\ref{Vsc}) gives the quantum contribution to the classical Hamiltonian in this semiclassical model.

It is not hard to work out the following expectation values:
\bea
&&\langle \Psi | Q |\Psi\rangle =x, \quad\quad  \langle \Psi | P |\Psi\rangle=p, \n\\
&&\langle \Psi |(Q-x)^2|\Psi\rangle=\hbar G, \label{EV}\\
&&\langle \Psi |(P-p)^2|\Psi\rangle=4\hbar\,G\Pi^2 + \frac{\hbar}{4G}.\n
\eea
It is clear that $x$ and $p$ are the expectation values of the operators $Q$
and $P$, respectively.  Also, $\hbar G$ is the mean fluctuation of the position
and that $G>0$.  $\Pi$ is related to the mean fluctuation of $P$.
The uncertainty relation is 
\be
\Delta Q\,\Delta P = \frac{\hbar}{2} \sqrt{1 + (4 G\Pi)^2}.
\label{UR}
\ee

Varying $\Gamma$ with respect to  $x,~p,~G$ and $\Pi$ gives the
Hamilton equations
of motion 
\bea
&&\dot{x}=\frac{p}{M},  \quad\quad  {\dot G}=\frac{4}{M} G\Pi, 
\n\\
&&\dot{p}=-V^\prime -\frac{\hbar}{2} G V^{\prime\prime\prime},
\label{qEOM}\\
&& \dot{\Pi}=\frac{1}{8M}G^{-2} - \frac{2}{M}\Pi^2 -\frac12 V^{\prime\prime}.\n
\eea

%-------------------------

\section{ Effective potential and semi-classical Kramers escape rate}

To derive  the effective potential appropriate for our model of the semiclassical Kramers problem, we make the following assumptions:
1) the quantum fluctuations should be as small as possible,  so that the system is as close to the classical limit as possible;
2) in the strong damping limit, it is reasonable to assume the particle stays mainly in the bottom of the well before crossing the barrier.

The requirement in assumption (1) can be attained by requiring that the uncertainly relation between $Q$ and $P$ be minimal.  This requires $\Pi=0$ from (\ref{UR}).  Note that we must have $G>0$ in order that  $V_{sc}(x)$ in (\ref{Vsc}) be a regular function.

Assumption (2) implies the particle is nearly in a quasi-stationary state at $x_0=0$. So we could assume that the state variables vary very slowly.  Here we consider the situation in which the quantum fluctuations are nearly time-independent, i.e., $\dot G=\dot\Pi=0$. 
That $\dot G=0$ is already implied by $\Pi=0$ from the second equation in (\ref{qEOM}). Likewise we demand $\Pi=0$ is constant in time, i.e. $\dot \Pi=0$.   The forth equation in (\ref{qEOM}) then fixes the value of $G$ to be
\be
G=\frac{1}{2\sqrt{MV^{\prime\prime}(x_0)}}.
\label{G}
\ee

Putting these values back into $H_{sc}$ then gives us the effective potential
\be
{\hat V}(x)=V(x)+\hbar\left[\frac{1}{8MG}  
+  \frac12 G V^{\prime\prime}\right],
\label{Veff}
\ee
where $G$ is now given by (\ref{G}).
Equation of motion of $p$ in (\ref{qEOM}) is the same as the Newton second law using  ${\hat V}(x)$.  

We now treat  the Kramers problem in the presence of both the thermal and the quantum fluctuations as simply the Kramers escape activated by thermal noise in the semiclassical potential ${\hat V}(x)$.  The derivation of the escape rate is then carried out in exactly the same way as in the classical case.  It is noted that, for strong damping and small quantum fluctuations assumed here, the original potential $V(x)$ is deformed only slightly to $\hat{V}(x)$, and that the quadratic approximation at the minimum and maximum of $\hat{V}(x)$ is still appropriate. Thus the quantum corrected escape rate is given by the expression (\ref{r_c}), with all relevant quantities related to $V(x)$ replaced by those related to $\hat{V}(x)$, 
\be
r_{sc} =\frac{\hat\omega_0\hat\omega_b}{2\pi\gamma}\,e^{-\beta [\hat{V}(\hat{x}_b)-\hat{V}(\hat{x}_0)]}.
\ee
Here $\hat{x}_0$ and $\hat{x}_b$ are the positions of the minimum and maximum of $\hat{V}(x)$, and 
$\hat\omega_0\equiv \sqrt{\hat{V}^{\prime\prime}(\hat{x}_0)/M}$ and $\hat\omega_b\equiv \sqrt{|\hat{V}^{\prime\prime}(\hat{x}_b)|/M}$.

Again, under the approximation adopted here, we can set
$\hat{x}_{0,b}\approx x_{0,b}$ and $\hat\omega_{0,b}\approx \omega_{0,b}$. With these values, we finally have
\be
r_{sc}\approx r_c e^{\frac12 \beta\hbar G \left[V^{\prime\prime}(x_0) +|V^{\prime\prime}(x_b)|\right]},
\label{r_sc}
\ee
where $r_c$ is given by (\ref{r_c}).
The second exponential factor represents the enhancement of the escape rate due to quantum fluctuations.
Eq.\,(\ref{r_sc}) can be expressed in terms of $\omega_0$ and $\omega_b$ as
\be
r_{sc}\approx r_c \exp\left\{\frac{\hbar \beta\left(\omega_0^2+ \omega_b^2\right)}{4\omega_0}\right\}.
\label{r}
\ee

\section{Discussion}

The result (\ref{r_sc}) is consistent in form with those from the quantum Smoluchowski equations considered in \cite{QK1,A3}, except the quantum fluctuation  $\hbar G$ is replaced by the symbol $\lambda$ in these two references. However, the viewpoints adopted in the derivations of the results are different. 

In our semiclassical approach, quantum corrections are incorporated  in  the drift potential $V(x)$, and their values are determined by quasi-stationary conditions and minimal uncertainty relation.  On the contrary, in \cite{QK1,A3}
a quantum version of the Smoluchowski equation was deduced heuristically 
by modifying the diffusion coefficient $D=k_B T/M\gamma$ to \cite{S1,S2}
\be
D_q=D \left[1-\frac{\lambda V^{\prime\prime}(x)}{k_B T}\right]^{-1},
\ee
where $\lambda=\langle x^2 \rangle - \langle x^2 \rangle_{cl}$ is the quantum fluctuation in position. The values of $\lambda$ is substantiated by path-integral computations for different damping cases and temperature regimes.  For instance, in the strong damping limit considered in this work, the escape rate by this approach is determined to be \cite{QK1}
\be
r_q\approx r_c \exp\left\{\frac{\hbar \beta\left(\omega_0^2+ \omega_b^2\right)}{2\pi\gamma}\left[\psi\left(1+\frac{\hbar\beta\gamma}{2\pi}\right) -\psi(1)\right]\right\},
\ee
where $\psi(z)$ is the digamma function.
This is similar to (\ref{r}), except in the exponent the factor $\omega_0$ is replaced by $\gamma$ and a numerical factor in terms of the digamma function. The difference is of course due to the different assumptions employed in the two approaches.  Computing quantum corrections by the path integral method has the advantage that it could cover different damping regimes and for non-Markovian noise as well, but the results depend very much on the assumed frequency distribution of the bath of oscillators  that coupled to the Brownian particle, and continuation into imaginary time is used, which is physically less intuitive.   Also, as mentioned before, the form of the  quantum modified drift coefficient in this quantum Smoluchowski equation is assumed heuristically.   In contrast, the semiclassical approach adopted here is more easily understood in terms of classical picture, and the quantum corrected drift potential (not the diffusion coefficient), is derived by a simple variational principle.  
In this respect, we note that in an earlier version of the quantum Smoluchowski equation proposed in \cite{A1,A2,S1,S2},  quantum corrections appeared in both the drift potential and the diffusion term.  As a result, the exponent in the quantum enhancement factor in (\ref{r_sc}) is doubled. This has been noted in \cite{W2,QK1,A3}.

In this work we have considered the stochastic force on the Brownian particle to be the Gaussian noise.  It should be of interest to extend the work to cases with colored noise. 

%%%%%%%%%%

%%%%%%%%

%\end{document}

\section*{Acknowledgments}

The work is supported in part by the Ministry of Science and Technology (MOST)
of the Republic of China under Grants   NSTC 112-2112-M-032-007 and NSTC 113-2112-M-032-010.

%%%%%%%%%%

%----------------------
\end{document}